\documentclass[twocolumn,secnumarabic,amssymb, nobibnotes, aps, prd,
 superscriptaddress]{revtex4}

\usepackage[utf8]{inputenc}
\usepackage{graphics,graphicx}
\usepackage{amsmath,amssymb,amsfonts,latexsym,cancel}
\usepackage{bm}

\begin{document}
\title{Incompressible relativistic spheres: Electrically charged stars,
compactness bounds, and quasiblack hole 
configurations}


\author{Jos\'e D. V. Arba\~nil}\email{dfi6jdav@joinville.udesc.br}
\affiliation{Centro
de Ci\^encias Naturais e Humanas, Universidade Federal do ABC, Rua
Santa Ad\'elia 166, 09210-170 Santo Andr\'e, SP, Brazil.}
\affiliation{Centro
de Ci\^encias Tecnol\'ogicas, Universidade do Estado de Santa Catarina, Rua
Paulo Malschitzki S$/$N, 89219-710 Joinville, SC, Brazil.}
\author{Jos\'e P. S. Lemos}\email{joselemos@ist.utl.pt}
\affiliation{Centro Multidisciplinar de Astrof\'isica - CENTRA,
Departamento de F\'{\i}sica, Instituto Superior T\'ecnico - IST,
Universidade de Lisboa - UL, Avenida Rovisco Pais 1, 1049-001
Lisboa, Portugal.}
\author{Vilson T. Zanchin}\email{zanchin@ufabc.edu.br}
\affiliation{Centro
de Ci\^encias Naturais e Humanas, Universidade Federal do ABC, Rua
Santa Ad\'elia 166, 09210-170 Santo Andr\'e, SP, Brazil.}

\begin{abstract} 
We investigate the properties of relativistic star spheres made of an 
electrically charged incompressible fluid, generalizing, thus, the 
Schwarzschild interior solution.  The investigation is carried by 
integrating numerically the hydrostatic equilibrium equation, i.e., the 
Tolman-Oppenheimer-Volkoff (TOV) equation, with the hypothesis that the 
charge distribution is proportional to the energy density. We  match the 
interior to a Reissner-Nordstr\"om exterior, and study some features of 
these star spheres such as the total mass $M$, the radius $R$, and the total 
charge $Q$. We also display the pressure profile. For star spheres made of a 
perfect fluid there is the Buchdahl bound, $R/M\geq 9/4$, a compactness 
bound found  from generic principles. For the Schwarzschild interior 
solution there is also the known compactness limit, the interior 
Schwarzschild limit where the configurations attain infinite central 
pressure, given by $R/M=9/4$, yielding an instance where the Buchdahl bound 
is saturated. We study this limit of infinite central pressure for the 
electrically charged stars and compare it with the Buchdahl-Andr\'easson 
bound, a limit that, like the Buchdahl bound for the uncharged case, is 
obtained by imposing some generic physical conditions on charged 
configurations.  We show that the electrical interior Schwarzschild limit of 
all but two configurations is always below the Buchdahl-Andr\'easson limit, 
i.e., we find that the electrical interior Schwarzschild limit does not 
generically saturate the Buchdahl-Andr\'easson bound. We also find that the 
quasiblack hole limit, i.e., the extremal most compact limit for charged 
incompressible stars, is reached when the matter is highly charged and the 
star's central pressure tends to infinity. This is one of the two instances 
where the Buchdahl-Andr\'easson bound is saturated, the other being the 
uncharged, interior Schwarzschild solution.

\end{abstract}
 \maketitle

\section {Introduction}
\label{sec-introd}

The search for matter field solutions in general relativity started
with Schwarzschild \cite{incom_schwarzschild} who studied a perfect
fluid with spherically symmetric energy density distribution given by
$\rho(r)$ and pressure $p(r)$, where $r$ is the radial coordinate, for
a constant energy equation of state, $\rho=\rho_0$, with $\rho_0$ a
constant, matched to a static spherically symmetric vacuum exterior
spacetime.  The exterior solution is the Schwarzschild solution, while
the whole solution, comprised of the interior plus exterior, is called
the Schwarzschild interior solution \cite{incom_schwarzschild}.
Schwarzschild also obtained a compactness limit, the interior
Schwarzschild limit. He showed \cite{incom_schwarzschild} that at
$R/M=9/4$ the central pressure becomes infinite, and for smaller $R$
there is no central pressure that can sustain the configuration, and
it presumably collapses, where $R$ is the radius of the configuration
and $M$ its mass.  Tolman \cite{tolman} and Oppenheimer and Volkoff
\cite{oppievolkoff} then discussed in detail the structure of those
matter field equations by writing the pressure gradient field in a
convenient form, now called the Tolman-Oppenheimer-Volkoff (TOV)
equation.  Volkoff \cite{volkoff} and then Misner in his lectures
\cite{misner} showed how the Schwarzschild interior solution can be
extracted directly from the TOV equation and rederived the limit of
$R/M=9/4$ as had been found by Schwarzschild
\cite{incom_schwarzschild}.  Nowadays, standard texts in general
relativity exhibit the Schwarzschild interior solution and its limit.
A constant density equation of state for a fluid in a star is
certainly interesting not only from a historical perspective.
Notably, a constant density is achieved when the matter is negligibly
compressible and can sustain huge pressures. This happens when the
velocities of the matter particles become relativistic, i.e., when the
temperatures are of the order of the particles' rest mass if the
matter is composed of bosons, or when the Fermi levels are again of
the order of the particles' rest mass if the matter is composed of
fermions.  This means that in both cases the densities approach one
particle per cubic Compton wavelength.  In addition, it is from an
incompressible equation of state that one extracts the important
compactness Schwarzschild limit, which can then be compared with other
compactness bounds, and the results give a clean and robust picture.
Of course, an incompressible equation of state also has some
drawbacks, as the speed of sound through such a medium is infinite,
although this does not have a major influence in the overall structure.
There are many other equations of state that can be used to study
compact stars.  Indeed, for instance, Oppenheimer and Volkoff
\cite{oppievolkoff} applied the method to study compact neutron stars,
stars that do not obey an incompressible equation of state. As
expected then, the most compact neutron stars are not as compact as
the interior Schwarzschild limit allows.

Buchdahl \cite{buchdahl} took further interest in general relativistic
equilibrium and proved that, under certain conditions, a spherically
symmetric matter configuration could only exist when the radius $R$ of
the configuration divided by its mass $M$ satisfies the relation
$R/M\geq9/4$. The result is quite general. Indeed the only assumptions
for providing this result are that the fluid's energy density $\rho$
is non-negative and decreasing outward, 
and the pressure $p$ is non-negative
and isotropic, i.e., the fluid is perfect, and the boundary defined by
$p=0$ is matched to the exterior Schwarzschild solution. The
inequality is a compactness bound, the Buchdahl bound, and the value
$R/M=9/4$ is the Buchdahl limit. It is a limit of limits.  It also
means a star made of such a fluid cannot approach its own
gravitational radius $R/M=2$ and so trapped surfaces are eliminated.
Although derived from completely different means, the interior
Schwarzschild limit (i.e., the limit of infinite central pressure for
incompressible spheres) \cite{incom_schwarzschild} and the Buchdahl
limit (i.e., the limit for perfect fluids obeying reasonable physical
conditions) \cite{buchdahl} coincide, and so the interior
Schwarzschild solution with infinite pressure saturates the Buchdahl
bound and, thus, realizes the Buchdahl limit.

The interest of adding electric charge to the matter in writing the
TOV equation was understood by Bekenstein in \cite{bekenstein}.
Indeed, although a significant excess of electric charge in stars is
unlikely, its study in stars is of great interest since it mimics
other fields and possible alterations in the gravitational field as
proposed in alternative theories of gravity.  Perhaps, the most simple
electric star is a star with a Schwarzschild incompressible interior
with some type of added electric charge distribution.  Adding electric
charge to these configurations can bring some insight to their overall
structure in more complex situations.  Thus, in this spirit and
sticking to incompressible fluids, de Felice and collaborators
\cite{defelice_siming,defelice_yu} studied the structure of electric
stars with a power-law charge distribution $q(r)$ of the form
$q(r)=Q\,(r/R)^n$, for some exponent $n\geq3$, total charge $Q$, and
star's radius $R$, matching the solutions, at the star's boundary, to
the Reissner-Nordstr\"om solution for the given star's mass $M$,
radius $R$, and charge $Q$. Numerical techniques were used. They also
analyzed the most compact stars for each $Q$. Anninos and Rothman
\cite{annroth} also studied incompressible fluids with somewhat more
intricate charge distributions. Now, in the case the radius $R$ of the
charged star is taken to the gravitational radius of the star itself,
the configuration is called a quasiblack hole. It is an extremal
configuration.  In this extremal configuration the total electric
charge of the star $Q$ is equal to the star's mass $M$, $Q=M$, and the
configuration has a radius $R$ equal to its own gravitational radius
$r_+=M$, i.e., $R=M$.  This configuration is highly compact and is a
limit for the case $Q=M$. In
\cite{defelice_siming,defelice_yu,annroth} such configurations were
detected. The stability of physical system, in particular of
electrically charged stars, is always of importance.  An analysis of
the stability of such stars against radial perturbations has been
performed in \cite{defelice_siming,annroth} with the conclusion that
for a range of the star parameters, and depending on how stiff the
perturbed matter is, these systems can be stable.

There has also been interest in finding a Buchdahl electric limit
formula which started in the works \cite{sim,bohmerharko}.  In
\cite{giulianirothman}, among other cases studied, the $\rho=\rho_0$
solution, with $\rho_0$ a constant, with the $n=3$ power-law
distribution for the charge given in
\cite{defelice_siming,defelice_yu}, was
revisited with the aim of finding the most compact configuration and a
Buchdahl electric limit formula. Up to now, the sharpest bound has
been given by Andr\'easson \cite{andreasson_charged}. This
Buchdahl-Andr\'easson bound for electrically charged matter is taken
from the condition $p+2p_T\leq\rho$, where $p$ and $p_T$ are the
radial and tangential pressures, respectively, and some other
reasonable physical conditions \cite{andreasson_charged}.  It is given
by $R/M\geq9/\left(1+\sqrt{1+3\,Q^2/R^2}\right)^{2}$.  Surprisingly,
although derived from completely different hypotheses, the Buchdahl
bound for the uncharged $Q=0$ case $R/M\geq9/4$ is reobtained. It
also gives the quasiblack hole limit, for $Q=R$ one obtains $R=M$, and
so $Q=M$ also. Charged shells saturate the Buchdahl-Andr\'easson bound
\cite{andreasson_charged}, whereas the $\rho=\rho_0$ solution, with
$\rho_0$ a constant, and $q(r)=Q\,(r/R)^3$ of \cite{giulianirothman}
do not, as noticed in \cite{andreasson_charged}. Presumably, the whole
set of solutions found in \cite{defelice_siming,defelice_yu} also
does not saturate generically the Buchdahl-Andr\'easson bound.

Motivated by these settings we want to further investigate
relativistic stars made of an incompressible fluid, $\rho=\rho_0$,
with $\rho_0$ being a constant.  In order to probe the genericity of
the results mentioned above we will assume a new charge distribution
$q(r)$, namely, we take the electric charge density $\rho_e$ as
proportional to the energy density, $\rho_e=\alpha\rho_0$, with
$\alpha$ a number obeying $0\leq\alpha\leq1$. This is a possible
assumption as one might think that the net charge is inherent to each
particle, having arisen from some physical consorted process. Then the
charged distribution $q(r)$ is taken from the Maxwell field equation
$\frac{dq(r)}{dr}=4\pi\rho_{e}(r)\,r^{2} \sqrt{A(r)}$, where $A(r)$ is
the $g_{rr}(r)$ component of the spacetime metric. This gives a $q(r)$
different from the ones considered in
\cite{defelice_siming,defelice_yu,annroth}, namely, $q(r)=Q\,(r/R)^n$
(see also \cite{giulianirothman}). Other $q(r)$ distributions can be
thought of, such as one that has most of the charge at or near the
core of the star. Here we stick to $\rho_e=\alpha\rho_0$.  We then
analyze the structure of such spheres and study the compactness
bounds, namely, we numerically probe the electric interior
Schwarzschild limit with an electric charge $Q$ within the range
$0\leq Q\leq M$ and compare the results to the Buchdahl-Andr\'easson
bound \cite{andreasson_charged}.  We also study the quasiblack hole
limit $Q=R=M$ of this set of configurations.

It is worth mentioning
that there are equations of state, other than an
incompressible one, that can be used in the study of electrically
charged stars and that should be mentioned.  For instance, Cooperstock
and de la Cruz \cite{cooperstock} and Florides \cite{florides} used an
equation of the form $\rho(r) + {q^2(r)}/{8\pi\,r^4}={\rm constant}$
and studied some properties of the solutions (see also
\cite{guilfoyle}).  Polytropic equations were used in
\cite{zhang,raymalheirolemoszanchin,ghezzi2005,siffert}, where star
configurations and their structure were studied, and the Schwarzschild
electric limit for the given equation of state and for a given charge
was considered.  There are also works that treat a quark deconfining
phase in compact neutron 
stars \cite{quarkdecof1}, where an unbalance of
electric or color charge might appear.  These works
\cite{cooperstock,florides,guilfoyle,zhang,raymalheirolemoszanchin,ghezzi2005,siffert,quarkdecof1}
stayed away from the extremal quasiblack hole configuration.  Works
that did consider the Schwarzschild electric limit and the quasiblack
hole limit are (i) works that analyzed the Cooperstock-de la
Cruz-Florides equation of state \cite{cooperstock,florides,guilfoyle}
as in \cite{lemosezanchin_QBH_pressure}, (ii) works that used a
generic polytropic equation of state with any exponent as in
\cite{alz-poli-qbh}, (iii) works with a dust equation of state, i.e.,
where the fluid is composed of purely extremal charged matter, also
called electrically counterpoised matter, where static equilibrium
configurations at the quasiblack hole limit, with $Q=M$ and $R=M$ were
found \cite{bonnor_wick,lemosweinberg}, (iv) works that in addition to
the electric charge have a scalar charge \cite{bronn}, and of course
(v) works with an incompressible equation of state
\cite{defelice_siming,defelice_yu,annroth} as mentioned before, to
name a sample.  Note that some works, e.g. \cite{alz-poli-qbh}, have
been calling the Buchdahl limit what should in fact be named
the interior Schwarzschild limit.  Works that treat generically quasiblack
holes are \cite{lz1,lz4}, for a review see \cite{review}.

The paper is organized as follows. In Sec.~\ref{sec-basicequations}, we
give the general relativistic equations, the equations of structure
for a static spherically symmetric configuration, the equations of
state for the energy density and charge density, and discuss the
boundary conditions.  In Sec.~\ref{incom-spheres}, we study numerically
the structure of an electrically charged incompressible star. We give
the numerical input values and then find the mass of the relativistic
star as a function of the energy density, the radius as a function of
the energy density, and the charge as a function of the energy
density. The behavior of the mass, radius, and charge of the stars for
some central pressures is also shown.  In addition, the pressure
profile, i.e., $p(r)$, is displayed for some typical stars.  In
Sec.~\ref{sec-Buchdahl}, we present the electrical Schwarzschild limit
for these stars and compare with the Buchdahl-Andr\'easson limit. In
Sec.~\ref{qbh-section1}, we study in detail the quasiblack hole limit
of such a relativistic star and give the behavior of the redshift at
the surface of the quasiblack hole as a function of the star's
intrinsic parameters.  In Sec.~\ref{sec-conclusion}, we conclude. In
the Appendix we give the equations of structure in dimensionless form.

\section{General relativistic equations}
\label{sec-basicequations}

\subsection{Basic equations}

We are interested in analyzing the properties of highly compacted
charged spheres as described by Einstein-Maxwell equations
with charged matter (in this section we put $c=1$ and $G=1$),
i.e., 
\begin{eqnarray}
&& G_{\mu\nu}=8\pi T_{\mu\nu},\label{eqs de einstein}\\
&& \nabla_{\nu}F^{\mu\nu}=4\pi J^{\mu}, \label{eqs de maxwell}
\end{eqnarray} 
where greek indices are spacetime indices running from $0$ to $3$, with $0$ 
being a time index. 
The Einstein tensor $G_{\mu\nu}$ is defined in terms of the Ricci
tensor $R_{\mu\nu}$, the metric tensor $g_{\mu\nu}$, and
the Ricci scalar $R$ by the well-known relation 
$G_{\mu\nu} =R_{\mu\nu} -\frac{1}{2}g_{\mu\nu} R$. 
$T_{\mu\nu}$ stands for the
energy-momentum tensor, which in this study is written as a sum of two terms,
\begin{equation}\label{tensor1 de energia momento}
T_{\mu\nu}=E_{\mu\nu}+M_{\mu\nu}\,.
\end{equation}
The first part $E_{\mu\nu}$ is 
the electromagnetic energy-momentum tensor, which is given in terms of
the Faraday-Maxwell tensor $F_{\mu\nu}$ by the relation
\begin{equation}\label{tensor de energia momento}
E_{\mu\nu}=\frac{1}{4\pi}\left(
F_{\mu}\hspace{0.1mm}^{\gamma}F_{\nu\gamma}-
\frac{1}{4}g_{\mu\nu}
F_{\gamma\beta}F^{\gamma\beta}\right)\,.
\end{equation}
The matter energy-momentum content of the spacetime is
represented by $M_{\mu\nu}$, which has the form of
the energy-momentum tensor of a perfect fluid,
\begin{equation}
M_{\mu\nu}=(\rho+p)U_{\mu}U_{\nu}+ pg_{\mu\nu},
\end{equation}
with $\rho$ and $p$ being the energy density and the pressure of the
fluid, respectively, and $U_\mu$ is the fluid four-velocity.  
Equation~\eqref{eqs de maxwell} is the Maxwell equation,
stating the proportionality between the covariant 
derivative $\nabla_\nu$ of the Faraday-Maxwell tensor $F_{\mu\nu}$
and the electromagnetic four-current $J_\mu$. For a charged fluid, this 
current is given in terms of the electric charge density $\rho_e$ by
\begin{equation}
J^{\mu}=\rho_{e}U^{\mu}\,.
\end{equation}
The other Maxwell equation $\nabla_{[\alpha}F_{\beta\gamma]}=0$,
where $[...]$ means antisymmetrization, 
is automatically satisfied for a properly defined
$F_{\mu\nu}$.

\subsection{Equations of structure}
\label{sec-equilibriumequations}

The charged incompressible fluid spheres considered here are described by a
static fluid distribution with spherical symmetry, in such a way that the
line element is of the form
\begin{equation}\label{geral_metric}
ds^2=-B(r)dt^{2}+A(r)dr^2+r^2\left(d\theta^2+\sin^{2}\theta
d\phi^{2}\right),
\end{equation}
where $t$, $r$, $\theta$, and $\phi$ are the usual Schwarzschild-like
coordinates, and the metric potentials $A(r)$ and $B(r)$
are functions of the radial coordinate $r$ only.

The assumed spherical symmetry of the spacetime implies that the only
nonzero components of a purely electrical
Faraday-Maxwell tensor $F^{\mu\nu}$ are
$F^{tr}=-F^{rt}$, where $F^{tr}$ is a function of the radial
coordinate $r$ alone, $F^{tr}=F^{tr}(r)$. The other components of
$F^{\mu\nu}$ are identically zero. Hence, the only nonvanishing
component of the Maxwell equation (\ref{eqs de maxwell}) is
given by 
\begin{equation}
\label{continuidadedacarga}
\frac{dq(r)}{dr}=4\pi\rho_{e}(r)\,r^{2} \sqrt{A(r)},
\end{equation}
with $q(r)=r^2\sqrt{A(r)\,B(r)}\,F^{tr}(r)$ representing the total electric
charge inside a spherical surface labeled by the radial coordinate whose
value is $r$. With this, and from the line element
(\ref{geral_metric}), the Einstein equation (\ref{eqs de einstein})
yields the following relevant relations
\begin{eqnarray}
\frac{1}{A(r)}\left[1-\frac{r}{A(r)}\frac{dA(r)}{dr}\right]&=&1- 
8\pi r^2\left[\rho(r)
+\frac{q^{2}(r)}{8\pi r^{4}}\right], \hskip 0.6cm \label{G00final12}\\
\frac{1}{A(r)}\left[1+\frac{r}{B(r)}\frac{dB(r)}{dr}\right]&=&{1}
+ 8\pi r^2\left[p(r)-\frac{q^{2}(r)}{8\pi r^{4}}\right].\hskip 0.6cm
\label{G11final12}
\end{eqnarray}
Defining the new quantity $m(r)$ in such a way that
\begin{equation}\label{funcion metrica}
\frac{1}{A(r)}=1-\frac{2m(r)}{r}+\frac{q^{2}(r)}{r^{2}}, 
\end{equation}
and replacing $A(r)$ from Eq.~(\ref{funcion metrica}) into
Eq.~(\ref{G00final12}) it gives
\begin{equation}\label{continuidaddamassa}
\frac{dm(r)}{dr}=4\pi\rho(r) r^{2}+\frac{q(r)}{r}
\left[\frac{dq(r)}{dr}\right]\,.
\end{equation}
The new function $m(r)$ represents
the gravitational mass inside the sphere of radial coordinate $r$, and 
Eq.~\eqref{continuidaddamassa} then represents the energy conservation,
as measured in the star's frame. 

An additional equation is obtained from the 
contracted Bianchi identity
$\nabla_{\mu}T^{\mu\nu}=0$, which gives
\begin{equation}\label{conservacion2}
\frac{dB(r)}{dr}=\frac{B(r)}{p(r)+\rho(r)}\left[\frac{q(r)}{2\pi
r^{4}}\frac{dq(r)}{dr}-2\frac{dp(r)}{dr}\right].
\end{equation}
Finally, replacing Eq.~(\ref{continuidadedacarga}) and the
conservation equation (\ref{conservacion2}) into 
Eq.~(\ref{G11final12}) it yields
\begin{equation}\label{tov}
\frac{dp}{dr}=-(p+\rho)\frac{\left[4\pi pr+\dfrac{m}{r^{2}}
-\dfrac{q^{2}}{r^{3}}\right]}{\left[1-\dfrac{2m}{r}
+\dfrac{q^{2}}{r^{2}}\right]}+\rho_{e}\sqrt{A\,}\,\frac{q}{r^{2} } ,
\end{equation}
where to simplify the notation we 
have dropped the functional dependence, i.e., 
$A(r)=A$, $m(r)=m$, $q(r)=q$, $\rho(r)=\rho$, 
$p(r)=p$, and $\rho_{e}(r)=\rho_{e}$. 
Equation (\ref{tov}) is the TOV
equation \cite{oppievolkoff,tolman},  
modified by the inclusion of electric
charge \cite{bekenstein} (see also \cite{alz-poli-qbh}).

\subsection{Equation of state and the charge density profile}
\label{EOS_charge_DP}

In the present model there are six unknown functions, $B(r)$, $m(r)$,
$q(r)$, $\rho(r)$, $p(r)$, and $\rho_{e}(r)$, and just four equations,
Eqs.~(\ref{continuidadedacarga}), (\ref{continuidaddamassa}),
(\ref{conservacion2}), and (\ref{tov}). Additional relations are obtained
from a model for the matter, i.e., from a model for the cold fluid, which
furnishes relations between the pressure and the energy density.  For an
electrically charged fluid, a relation defining the electric charge
distribution is also needed.

For the present analysis we assume an incompressible fluid, i.e.,
\begin{equation}\label{density_0}
\rho(r)=\rho_0\,,
\end{equation}
with $\rho_0={\rm constant}$.
So the energy density is constant along the whole star.

Following \cite{zhang,raymalheirolemoszanchin} (see also
\cite{siffert} and \cite{alz-poli-qbh}), we assume
a charge density proportional to the energy density,
 \begin{equation}\label{densicarga_densimassa}
\rho_e=\alpha\rho_0,
\end{equation}
where, in geometric units, $\alpha$ is a dimensionless constant which
we call the charge fraction. The charge density along the whole star
is, thus, constant as well.

We have now four equations: Eqs.~(\ref{continuidadedacarga}),
(\ref{continuidaddamassa}), (\ref{conservacion2}), and
(\ref{tov}); and four unknowns: $B(r)$, $m(r)$,
$q(r)$, and $p(r)$, as $\rho_0$ and
$\rho_e$ are given in 
(\ref{density_0}) and (\ref{densicarga_densimassa}), respectively.
The resulting set of
equations constitutes the complete set of structure equations which,
with some appropriate boundary conditions, can be solved
simultaneously.

\subsection{The boundary conditions and the exterior vacuum 
region to the star}
\label{BC}

The numerical integration of the system of equations is performed along
the radial coordinate $r$, from the center toward the surface of the star.
The conditions at the center of the star ($r=0$) are $m(r=0)=0$, $q(r=0)=0$,
$p(r=0) = p_c$,
$\rho(r=0)=\rho_{c}$, $\rho_e(r=0)=\rho_{ec}$, and
$A(r=0)=1$, where $p_c$ is the central pressure, and 
$\rho_{c}$ and $\rho_{ec}$ are the central energy density and the 
central charge distributions, respectively,
which by assumption are 
constant throughout the star.
The surface of the star is defined by the vanishing of the
pressure. Since the pressure decreases outwards, the integration is
stopped at the point $r=R$ for which $p(R)=0$. The solution is then matched
to the exterior Reissner-Nordstr\"om spacetime, with 
metric given by
\begin{equation}
ds^2 = - F(r) dT^2 + \dfrac{dr^2}{F(r)}+ r^2
\left(d\theta^2+\sin^{2}\theta
d\phi^{2}\right)\,,
\label{rnords1}
\end{equation}
where
\begin{equation}
F(r) = 1 -2M/r + Q^2/r^2\, ,
\label{rnords2}
\end{equation}
with the outer
time $T$ being proportional to the inner time $t$, $M$ and $Q$ being
the total mass and the total charge of the star,
respectively. The full set of boundary conditions at the surface of the star
is $B(R)=1/A(R)=F(R)$, $m(R)=M$, $q(R)=Q$, besides $p(R)=0$.

\subsection{Our aim}
\label{aim}

Using the whole set of
equations and boundary conditions, it is our interest to study the
structure of these incompressible charged stars.  Given values for
the  energy density $\rho_0$ and for the charge fraction $\alpha$, we can
analyze the system for different values of central pressure $p_c$. For
each $\alpha$, and thus for each $Q/M$, there is a set of star
solutions, each corresponding to a given $p_c$. The
limiting solution that has $p_c=\infty$ gives
the interior Schwarzschild 
limit. We then compare this interior Schwarzschild 
limit with the electric 
Buchdahl limit, or Buchdahl-Andr\'easson limit
\cite{andreasson_charged}, i.e., the 
minimum ratio for $R/M$ for a given $Q/M$.  
In addition, the
most extreme configuration, the one that has $Q/M=1$, has as the
limiting solution the quasiblack hole, i.e., a star whose boundary is
located at its own gravitational radius. We look in detail into
these solutions and to their corresponding features.

\section{The structure of a charged incompressible 
relativistic star sphere and the Buchdahl limit}\label{incom-spheres}

\subsection{Numerical input values}

We now analyze the structure of 
incompressible charged star spheres with $\rho_0={\rm constant}$ 
and $\rho_e=\alpha\rho_0= {\rm constant}$.
The set of coupled equations, 
Eqs.~(\ref{continuidadedacarga}), (\ref{continuidaddamassa}),
(\ref{conservacion2}), (\ref{tov}), 
\eqref{density_0}, 
and \eqref{densicarga_densimassa}, and the boundary conditions
adopted at the center are integrated up to the boundary 
of the star. The integration is performed 
upon putting the equations in a dimensionless form (see
the Appendix). For each given energy density $\rho_0$,
charge fraction $\alpha$ and central pressure $p_c$, the system of
equations is numerically solved using a fourth order Runge-Kutta
method.

Here we are going to analyze the numerical results and to plot a few graphs
showing several parameters of the stars, such as the radius, the gravitational
mass, the total charge, and some other interesting
quantities. In order to obtain the corresponding
values in appropriate units, it is convenient to
restore the gravitational constant $G$, while keeping the speed of light set
to unity, i.e., in this section, we use units such that $G=7.42611 \times
10^{-28}\,[{\rm m}/{\rm kg}]$ and $c=1$.

\begin{figure}[!ht]
\centering
\includegraphics[scale=1.15]{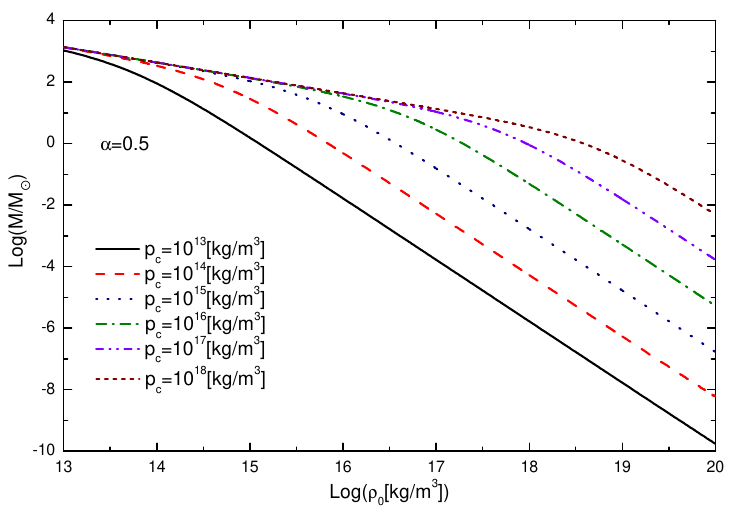}
\centering
\includegraphics[scale=1.16]{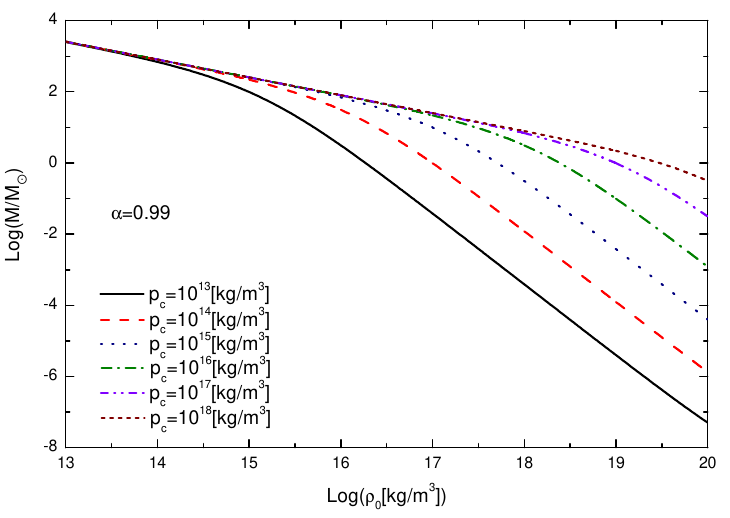}
\vspace*{-.3cm}
\caption{The ratio $M/M_\odot$ of the incompressible charged spheres as a
function of the energy density $\rho_0$ for six values of the central
pressure $p_c$ as indicated.
The top panel is for $\alpha=0.5$ and the bottom panel
for $\alpha=0.99$.}
\label{Log_rho_0vsLoM_Msol_dif_p_c}
\end{figure}

For ease of comparison with other works, we consider the energy density
and the central pressure in the ranges $1.0\times10^{13}[{\rm kg}/{\rm
m}^{3}] \leq\rho_0\leq1.0\times10^{20}[{\rm kg}/{\rm m}^{3}]$ and
$1.0\times10^{13}[{\rm kg}/{\rm m}^{3}] \leq
p_c\leq1.0\times10^{18}[{\rm kg}/{\rm m}^{3}]$, respectively. The
interval of energy densities considered here covers the range of values of
the central energy density of real neutron stars. On the other hand,
the values considered for the central pressures are within the range
of the central pressures considered in \cite{raymalheirolemoszanchin}.

Let us last mention that the charge fraction $\alpha$ is varied in
the interval $0\leq\alpha \leq 1$. However, it is important to say
that, in the intervals of energy density and central pressure
considered in the present work, the value  
$\alpha =0.99$ is in fact the largest value of the charge fraction
we have used in the numerical
analysis. This value was chosen for comparison to our previous work
\cite{alz-poli-qbh}. For larger values of $\alpha$ the
numerical calculations become very slow and eventually fail to
converge for $\alpha$ very close to unity. We have not found
equilibrium solutions for $\alpha > 1$.

The star's mass $M$, its radius $R$,  and total charge
$Q$, are found when the pressure at the surface of the object is equal
to zero $p(r=R)=0$. We present the main features of these three
quantities in the following subsections.

\subsection{Mass of a charged relativistic incompressible star as a function
of the energy density}

\begin{figure}[!ht]
\centering
\includegraphics[scale=1.16]{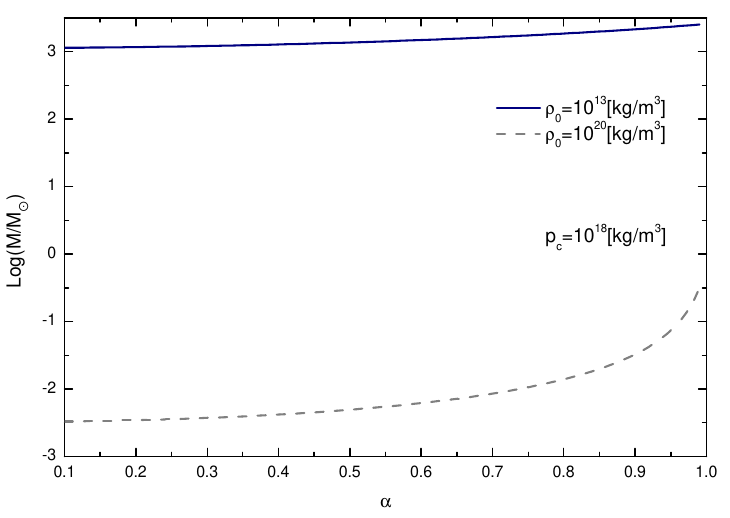}
\vspace*{-.3cm}
\caption{The ratio $M/M_\odot$ of the incompressible charged spheres
as a function of the charge fraction $\alpha$ for the central pressure
$p_c=1.0\times10^{18}\,[{\rm kg}/{\rm m}^{3}]$ and 
two values of the energy density $\rho_0$, namely, 
$1.0\times10^{13}\,[{\rm kg}/{\rm m}^{3}]$ and $1.0\times10^{20}\,[{\rm
kg}/{\rm m}^{3}]$.}
\label{alphavsLoM_Msol_dif_p_c}
\end{figure}

The mass of the star (in units of solar masses $M_{\odot}$) as
a function of the energy density $\rho_0$ is shown
in Fig.~\ref{Log_rho_0vsLoM_Msol_dif_p_c} for some values of the
central pressure. The top panel gives the curves
for the charge fraction $\alpha=0.5$ and
the bottom panel for 
$\alpha=0.99$. The values of the energy density and central
pressure are in the ranges $1.0\times10^{13}$
to $1.0\times10^{20}$[kg/m$^3$] and $1.0\times10^{13}$
to $1.0\times10^{18}$[kg/m$^3$], respectively. In all the presented cases,
we can note that the mass of the star decreases monotonically
with the increase of the energy density, and in a larger rate for small
pressures than for large pressures.
This can be understood in the sense that stars with large
energy densities have to have less mass in order for
equilibrium to be maintained. In addition, we can see that, for
low densities, large changes in the central pressure $p_c$
do not imply large changes in the mass, on the contrary,
the mass changes very little. It signals the fact 
that, for low densities, the increase on the central pressure is very
sensitive to small increases in the star's mass.
Notice also that for high densities and small central
pressures the masses  of the related incompressible stars are very small
compared to the case of high densities and large central pressures.

Figure \ref{alphavsLoM_Msol_dif_p_c} shows the dependence of the
mass of the object in units of solar masses
as a function of the charge fraction $\alpha$ for two values of the
energy density $\rho_0$ as indicated. The general feature is that the mass
grows monotonically with $\alpha$, and faster for higher densities. 
Without going into great detail, let us comment on the change in the mass
of an incompressible charged sphere when the charge fraction
$\alpha$ changes from  $0.5$ to $0.99$. For $p_c=1.0\times10^{18}[{\rm
kg}/{\rm m}^{3}]$ and $\rho_0=1.0\times10^{13}[{\rm kg}/{\rm m}^{3}]$,
the increase in the mass is about $84.8029\%$,
whereas for the same $p_c$ but with $\rho_0=1.0\times10^{20}[{\rm kg}/{\rm
m}^{3}]$ the increase in the mass is about $6220.67\%$.
From these examples, it
can be deduced that for low densities, in comparison with large densities,
the mass of the object does not change considerably with the increment
of the charge fraction $\alpha$.

\subsection{Radius of a charged relativistic incompressible star as a
function of the energy density}

\begin{figure}[t]
\centering
\includegraphics[scale=1.16]{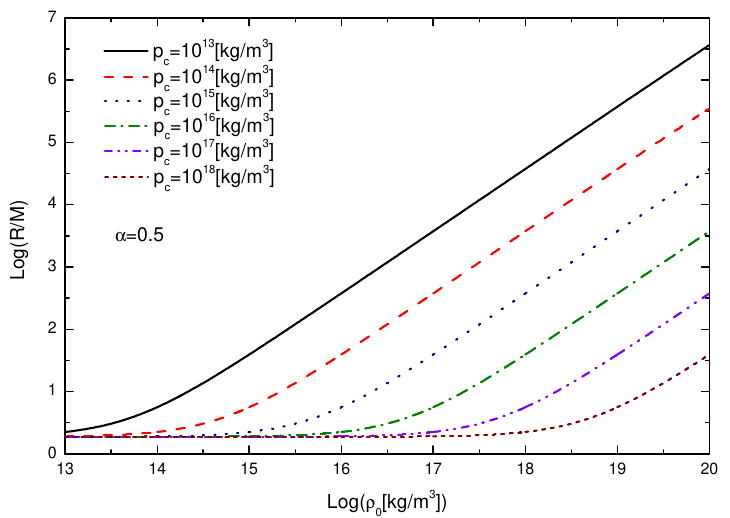}
\centering
\includegraphics[scale=1.15]{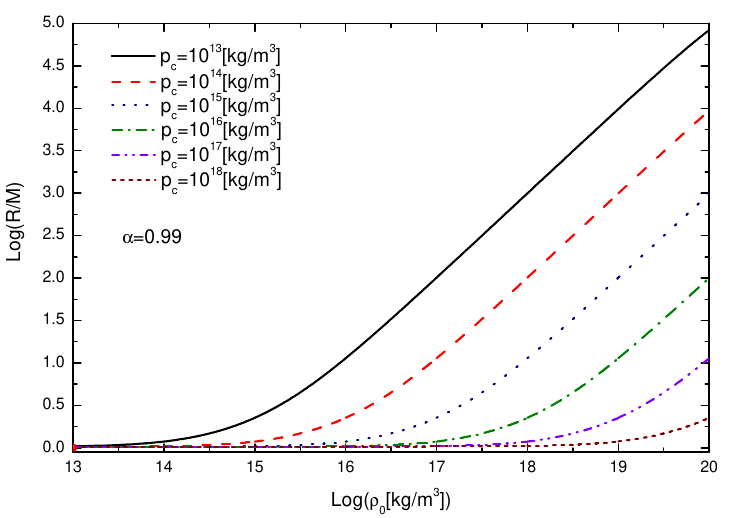}
\vspace*{-.3cm}
\caption{The ratio $R/M$ of the incompressible charged spheres as a function
of the energy density $\rho_0$ for six values of the central pressure
$p_c$,
as indicated. The top panel is for $\alpha=0.5$ and the bottom panel
for $\alpha=0.99$. The red point in the bottom panel indicates the
quasiblack hole configuration where $R/M\simeq 1.0$.}
\label{Log_rho_0vsLoR_M_dif_p_c}
\end{figure}

\begin{figure}[!ht]
\centering
\includegraphics[scale=1.15]{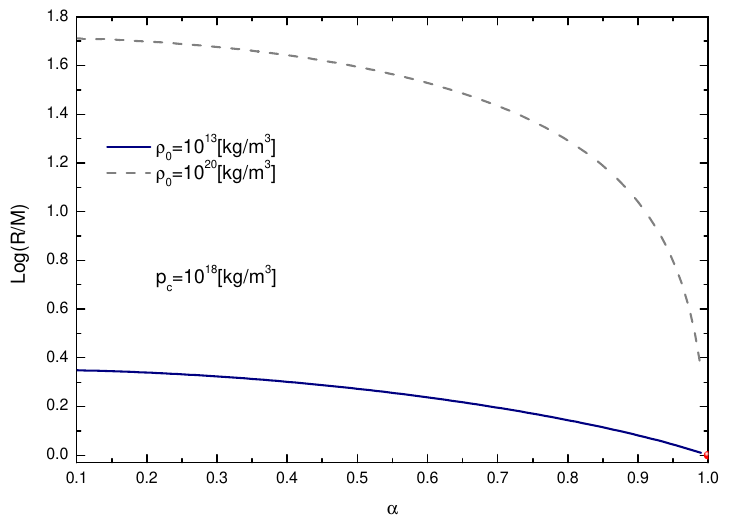}
\vspace*{-.3cm}
\caption{The ratio $R/M$ of the incompressible charged spheres as a function
of the charge fraction $\alpha$ for
a central pressure $p_c=1.0\times10^{18}[{\rm kg}/{\rm m}^{3}]$ and two 
values of the energy density, one case with $\rho_0= 1.0\times10^{20}[{\rm
kg}/{\rm m}^{3}]>p_c$, and the other case with $\rho_0= 1.0\times10^{12}[{\rm
kg}/{\rm m}^{3}]<< p_c$, as indicated. The red point
in the figure indicates the quasiblack
hole configuration where $R/M\simeq 1.0$.}
\label{alphavsLoR_M_dif_p_c}
\end{figure}

The radius of the star as a function of the energy density is shown in
Fig.~\ref{Log_rho_0vsLoR_M_dif_p_c}, where we plot the ratio $R/M$
against the energy density $\rho_0$ for some values of the central
pressure $p_c$. The energy density and central pressure values
in the figure are the same as
in Fig.~\ref{Log_rho_0vsLoM_Msol_dif_p_c}, so the
values of the considered energy densities are in the range
$1.0\times10^{13}$ to $1.0\times10^{20}$[kg/m$^3$].
Two values of the charge fraction were taken for comparison,
$\alpha = 0.5$ (top panel) and $\alpha =0.99$ (bottom panel). In all curves,
the relation $R/M$ grows with $\rho_0$, indicating that less compact
stars are found for large energy densities. On the other hand, we see
that there is also an influence of $p_c$ in the degree of compaction
of a star.  The larger $p_c$ is, the smaller the value of $R/M$. This
is of course expected, as the star gets more compact, the
gravitational action is stronger and there is a need for a greater
pressure throughout the star, including the central pressure $p_c$, to
counterbalance.

In Fig.~\ref{alphavsLoR_M_dif_p_c}, the dependence of 
$R/M$ is shown as a function of the charge fraction $\alpha$ for the
minimum and the maximum considered energy densities, i.e.,
$\rho_0=1.0\times10^{13}[{\rm kg}/{\rm m}^{3}]$ and
$\rho_0=1.0\times10^{20}[{\rm
kg}/{\rm m}^{3}]$, respectively, and for the central pressure
$p_c=1.0\times10^{18}[{\rm kg}/{\rm m}^{3}]$. The ratio $R/M$ decreases 
with the charge fraction, meaning that the stars bearing a large amount of
charge are more compact than the stars with little or no charge.
The case that has a central pressure $p_c$ much higher than the 
energy density $\rho_0$, the case shown in the full line,
is the case that approximates well the 
electric interior Schwarzschild limit for
each charge fraction $\alpha$. So, for 
$\alpha=0.0$, one gets $R/M\simeq 2.25$ which is 
the original interior Schwarzschild limit  $R/M=9/4$
\cite{incom_schwarzschild}.
For the case of $\alpha=0.99$, i.e., the extremal charge case,
we obtain $R/M\simeq 1.0$ which is in accord with the
limits set by Andr\'easson for charged spheres  
\cite{andreasson_charged}.
In this case the
limit is a quasiblack hole, the red point in the figure.

\subsection{Charge of a relativistic incompressible star as a function of the
energy density}

\begin{figure}[b]
\centering
\includegraphics[scale=1.15]{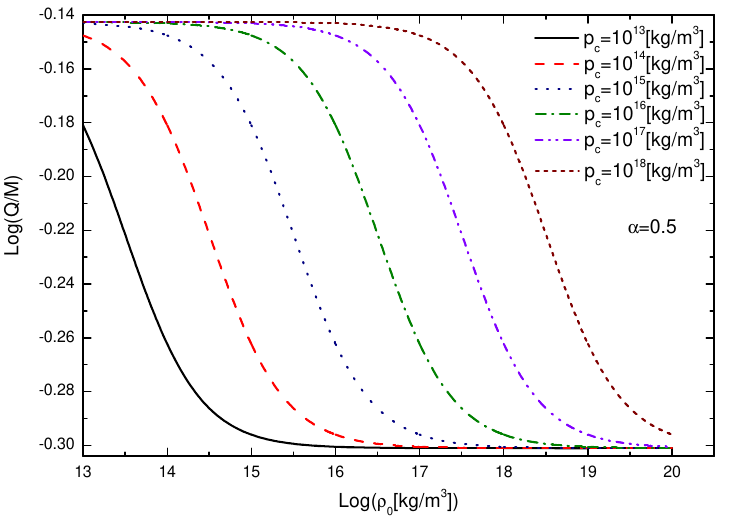}
\centering
\includegraphics[scale=1.155]{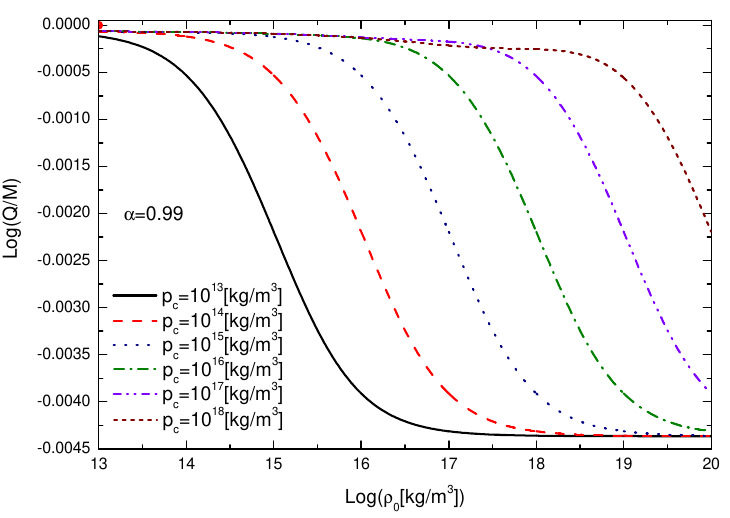}
\vspace*{-.3cm}
\caption{The ratio $Q/M$ of the incompressible charged spheres as a function
of the energy density $\rho_0$ for
six values of the central pressure $p_c$ as indicated. The top panel is for
$\alpha=0.5$ and the bottom panel
for $\alpha=0.99$. The red point in the figure at the bottom indicates
the quasiblack hole configuration where $Q/M\simeq 1.0$.}
\label{Log_rho_0vsLoQ_M_dif_p_c}
\end{figure}
\begin{figure}[!ht]
\centering
\includegraphics[scale=1.16]{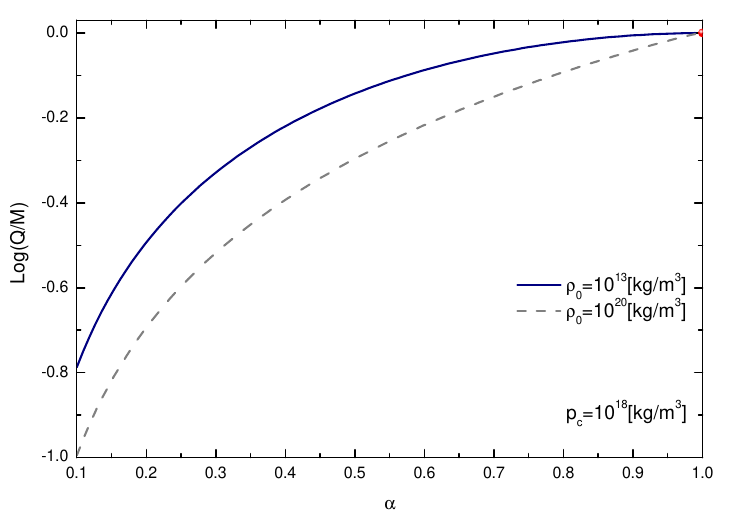}
\vspace*{-.3cm}
\caption{The ratio $Q/M$ of the incompressible charged spheres as a function
of the charge fraction $\alpha$ for
the central pressure $1.0\times10^{18}[{\rm kg}/{\rm m}^{3}]$ and two
values of the energy density as indicated. The red point indicates the
quasiblack hole limit where $Q/M\simeq 1.0$.}  \label{alphavsLoQ_M_dif_p_c}
\end{figure}

In Fig.~\ref{Log_rho_0vsLoQ_M_dif_p_c}, the charge to
mass relation $Q/M$ is plotted 
against the energy density $\rho_0$, for some
values of the central pressure, and for the charge fractions
$\alpha=0.5$ (top panel) and $\alpha=0.99$ (bottom panel). The red
point in the upper left corner of the bottom panel indicates the
quasiblack hole configuration ($Q/M =1.0$). It is seen that in all cases
the ratio $Q/M$ decreases as $\rho_0$ is increased. For sufficiently
small central pressures, each curve presents a plateau in the low
density region. The width of the plateau is larger (and higher) for
high charge fractions, and, as the energy density grows, the ratio
$Q/M$ decays very rapidly.  The minimum value of $Q/M$ is around the
corresponding value of $\alpha$. For instance, in the $\alpha =0.5$
case (top panel) the charge to mass ratio decreases from approximately
$Q/M\simeq 0.7$ at low energy densities to approximately $Q/M=0.5$ at
$\rho_0= 1.0\times 10^{20}[{\rm kg}/{\rm m}^{3}]$. This value is found for
$\rho_0\gtrsim 10^{3}p_c$. Another important feature is that the
fraction $Q/M$ increases with the central pressure. From this we see
that the larger the central pressure, the larger the amount of charge
the star admits. In fact, it is seen from the $\alpha = 0.99$ case
(bottom panel) that there are incompressible charged stars very close
to the quasiblack hole configuration, i.e., with $Q/M \simeq 1.0$.

The amount of charge supported by the incompressible spheres can also
be seen in Fig.~\ref{alphavsLoQ_M_dif_p_c}, which shows the
ratio $Q/M$ as a function of $\alpha$ for two values of the energy
density $\rho_0$. In this figure, the central pressure is
$1.0\times10^{18}[{\rm kg}/{\rm m}^{3}]$ and the energy densities are
$1.0\times10^{13}[{\rm kg}/{\rm m}^{3}]$ and $1.0\times10^{20}[{\rm
kg}/{\rm m}^{3}]$. The point in red in this figure, as in the bottom
panel of Fig.~\ref{Log_rho_0vsLoQ_M_dif_p_c}, indicates that for
$\alpha=1.0$ it is found $Q/M=1.0$.

\subsection{The interior pressure of a charged incompressible star}

\begin{figure}[!ht]
\centering
\includegraphics[scale=0.28]{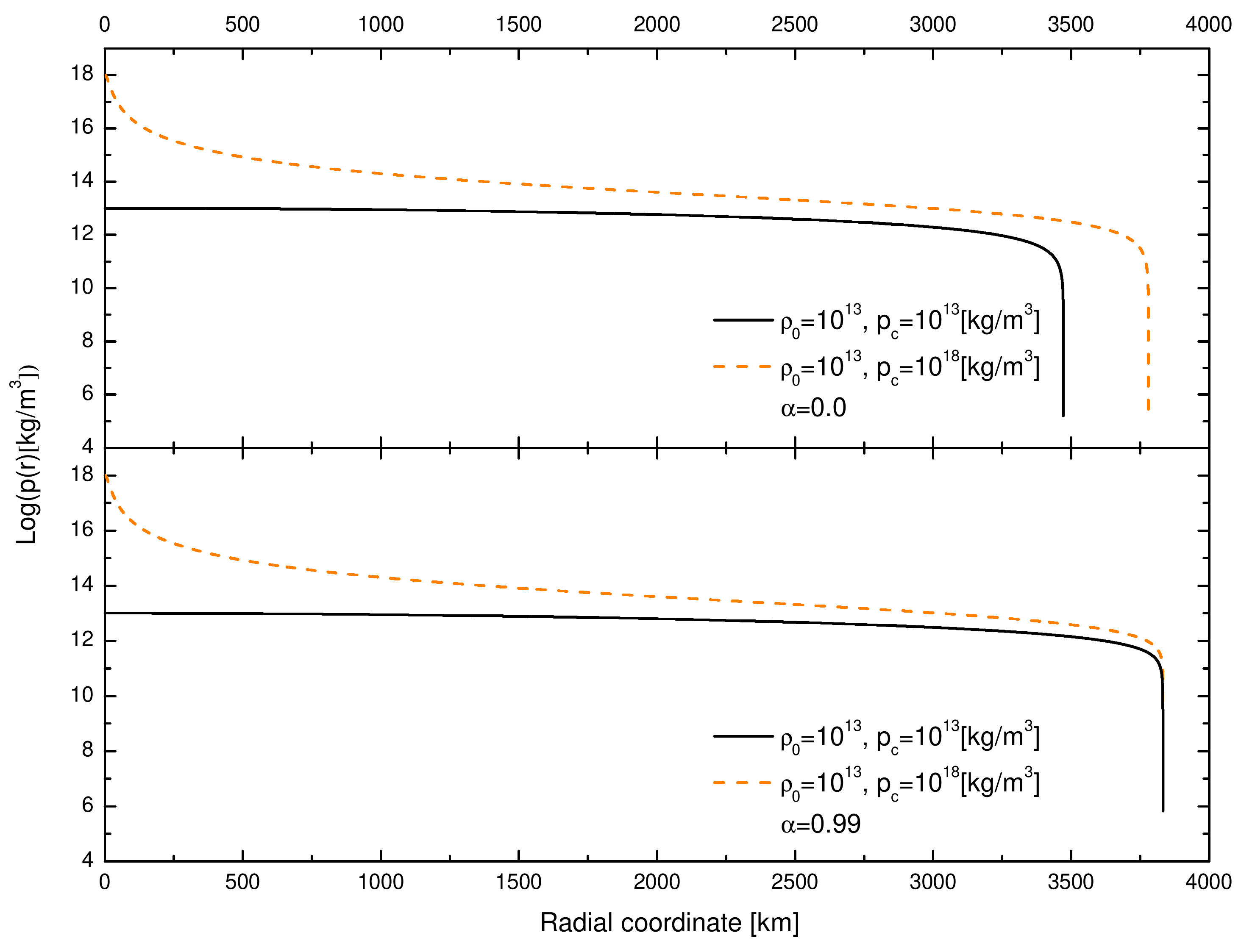}
\centering
\includegraphics[scale=0.28]{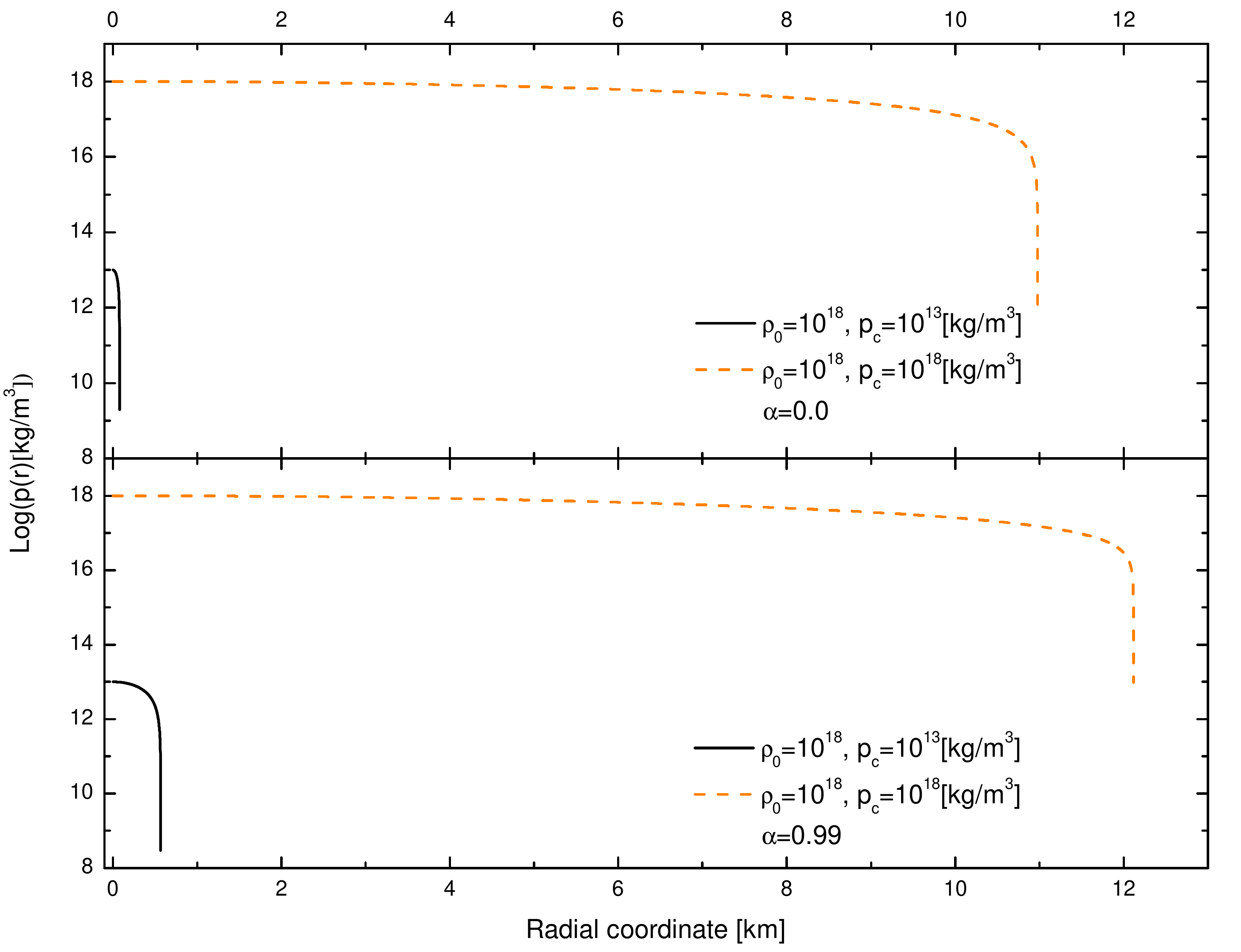}
\centering
\includegraphics[scale=0.28]{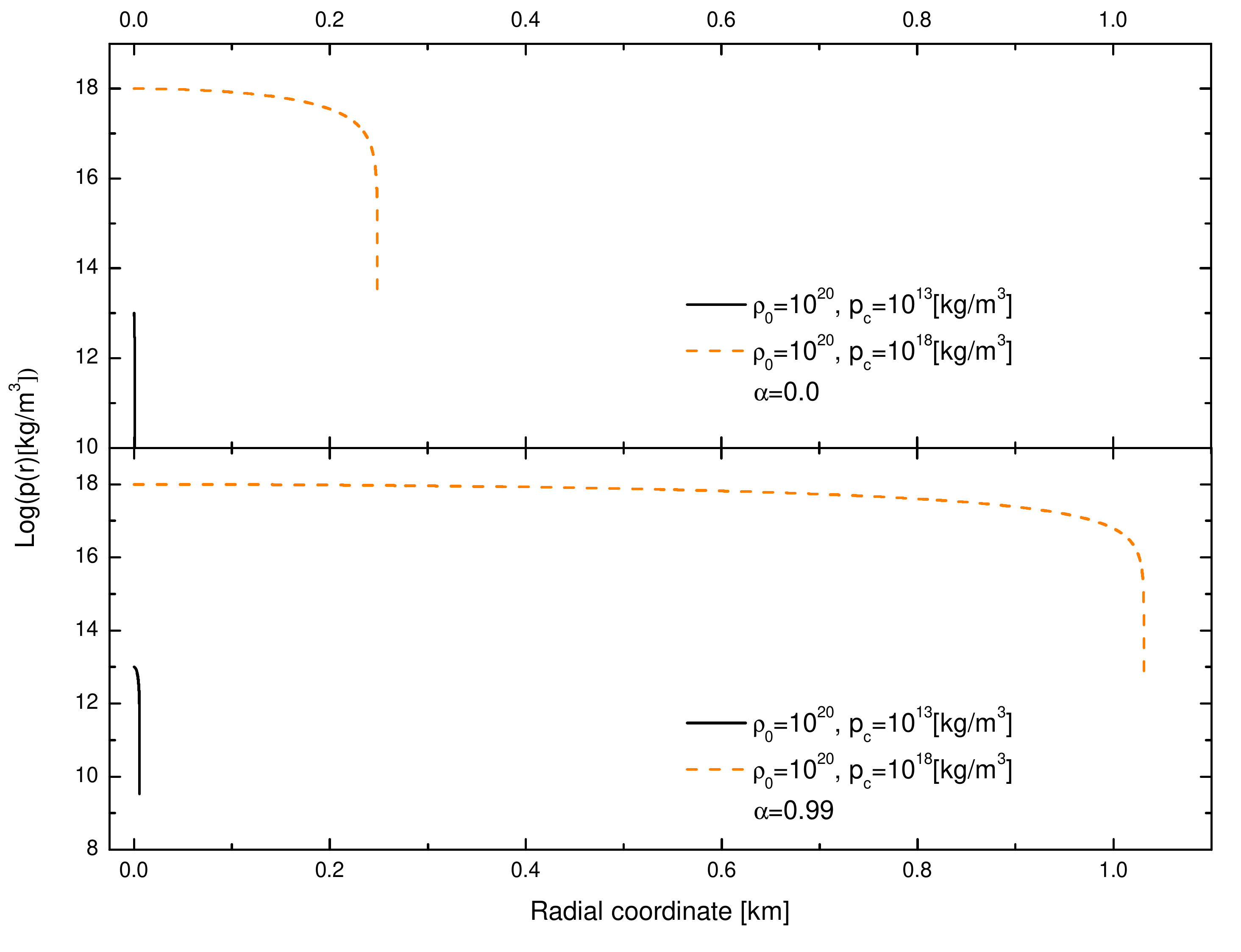}
\vspace*{-.3cm}
\caption{The figure shows the
pressure $p$ of the fluid as a function of the radial
coordinate $r$.  The panel at the top is for the energy density
$\rho_0=10^{13}[\rm kg/m^3]$, the one in the middle is for
$\rho_0=10^{18}[\rm kg/m^3]$, and the panel at the bottom is for
$\rho_0= 10^{20}[\rm kg/m^3]$.  Each panel displays $p(r)$ for two
values of charge fraction, $0.5$ and $0.99$ and for two central
pressures $p_c=10^{13}$ and $p_c=10^{18}[\rm kg/m^3]$.}
\label{rxp_rho}
\end{figure} 

We now study the behavior of the pressure inside these charged
incompressible spheres, i.e., we 
study $p(r)$.  Fi\-gu\-re \ref{rxp_rho} shows a
few particular cases of $p(r)$.  
The panel at the top is for the energy density
$\rho_0=10^{13}[\rm kg/m^3]$, the one in the middle is for
$\rho_0=10^{18}[\rm kg/m^3]$, and the panel at the bottom is for
$\rho_0= 10^{20}[\rm kg/m^3]$.  Each panel shows the results for two
values of the charge fraction, $\alpha =0.0$ and $\alpha = 0.99$, and
two central pressures $p_c=10^{13}[\rm kg/m^3]$ 
and $p_c=10^{18}[\rm kg/m^3]$.  As
expected, the pressure within the star decreases monotonically from
the center toward the surface of the object in all the analyzed
cases. For stars with central pressure smaller than the energy density
the pressure varies very slowly with the radial coordinate, going to
zero with a fast rate just close to the surface of the star.  The
general behavior is that of a star with constant pressure from the
center up to very close to the surface. On the other hand, stars with
central pressure larger than the energy density have a larger pressure
gradient. In particular, in the limit of a very large central pressure,
the pressure has to decrease abruptly with the radial coordinate to
reach vanishing values at the surface.  It is clearly observed that
for fixed energy density and charge fraction, the radius of a given
star increases with the central pressure.  In general, the
increase of the central pressure implies an increase of the
size $R$ of the equilibrium solutions and an increase in the mass $M$,
while the resulting star becomes more compact, in the sense that $R/M$
approaches unity.  From  Figs.~\ref{Log_rho_0vsLoR_M_dif_p_c} and
\ref{rxp_rho}, it is seen that more compact stars have larger
radii and larger mass and charge.

\section{The electric
interior Schwarzschild limit of a charged relativistic
incompressible star and the Buchdahl-Andr\'easson limit} 
\label{sec-Buchdahl}

\begin{figure}[!ht]
     \centering
     \includegraphics[scale=0.28]{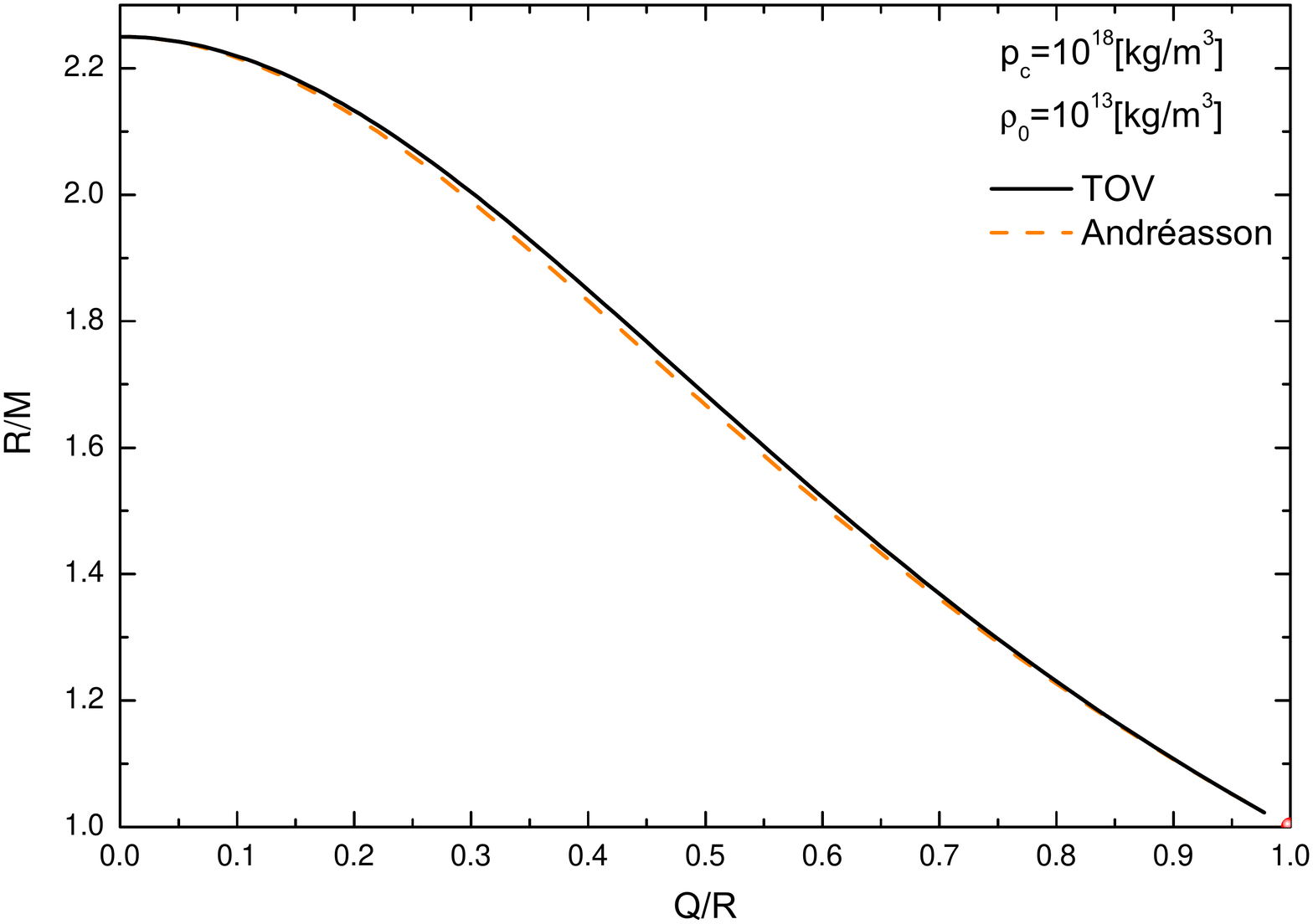}
		\vspace*{-.3cm}
\caption{The solid black line shows 
the electric interior Schwarzschild limit, i.e., 
the ratio $R/M$ versus the ratio $Q/R$ for
very large central pressures, here for  
$\rho_0=1.0\times10^{13}$[kg/m$^3$] and
$p_c=1.0\times10^{18}$[kg/m$^3$].  
The charge fraction varies from
$\alpha=0.0$ to $\alpha=1$ (numerically, the maximum
value of $\alpha$ we have managed
to achieve was $\alpha=0.99$). The line is found using 
the TOV equation for these specific stars. 
The red dashed line gives the Buchdahl-Andr\'easson limit, 
i.e., the values
found using the relation
$R/M=9/\left(1+\sqrt{1+3\,Q^2/R^2}\right)^{2}$.  
The two curves coincide in the boundary
points, namely, $Q/R=0$ and $Q/R=1$, but not in the 
intermediate points. The 
Buchdahl-Andr\'easson curve bounds the 
electric interior Schwarzschild limit curve
and is indeed a limit of limits. 
The point in red,  $Q/R=1$, represents the quasiblack
hole limit.}
     \label{Q_RvsR_M_TOV_andre_p_c_1E18_rho_1E13}
\end{figure}

The solid black line in
Fig.~\ref{Q_RvsR_M_TOV_andre_p_c_1E18_rho_1E13} gives the ratio $R/M$
as a function of $Q/R$ found by means of the TOV equation, 
 for the lowest energy density and the
largest central pressure used in the present analysis, i.e., for
$\rho_0=1.0\times10^{13}$[kg/m$^3$] and
$p_c=1.0\times10^{18}$[kg/m$^3$], respectively. Numerically, this
discrepancy of $10^5$ between $p_c$ and $\rho_0$ simulates the
electric interior Schwarzschild limit, as this happens for
$p_c/\rho_0\to\infty$.
Thus, the solid black line 
gives the electric interior Schwarzschild limit $p_c/\rho_0\to\infty$.
For $Q/R=0$, i.e., for the interior Schwarzschild solution, one gets
the original interior Schwarzschild limit of $R/M=9/4=2.25$
\cite{incom_schwarzschild}.  In the other extreme, for $Q/R=1$, one
gets the quasiblack hole $R/M=1$. In between these charges the values
of $R/M$ for this limit are displayed in the curve.
As noted before, numerically we work with $\alpha=0.99$
rather than $\alpha=1$, and so our results give rather
$Q/M\simeq 1.0$, $Q/R\simeq 1.0$, and 
$R/M\simeq 1.0$.

One should also compare the electric interior Schwarzschild
limit with the Buchdahl-Andr\'easson limit.
The Buchdahl limit \cite{buchdahl}, a limit of limits, is obtained by
imposing that the density of the star should be nonincreasing with
radius $r$, the pressure is isotropic, and a few other reasonable
assumptions.  For stars with nonzero electrical charge
there is the  Buchdahl-Andr\'easson limit, a generalized
Buchdahl limit found by
Andr\'easson \cite{andreasson_charged}
by imposing other conditions, such as 
$p+2p_T\leq\rho$, where $p$ and $p_T$ are the 
radial and tangential pressures, respectively, together with 
some other reasonable physical conditions. 
The Buchdahl-Andr\'easson bound is 
\cite{andreasson_charged})
\begin{equation}
\frac{R}{M}\geq\frac{9}{\left(1+\sqrt{1+3Q^{2}/R^{2}}\right)^{2}}\,,
\label{balimit}
\end{equation}
where the inequality gives the Buchdahl-Andr\'easson limit.
For $Q=0$, Eq.~(\ref{balimit}) gives the Buchdahl bound 
$R/M\geq 9/4$, of course.
In
Fig.~\ref{Q_RvsR_M_TOV_andre_p_c_1E18_rho_1E13} the red dashed line is
the Buchdahl-Andr\'easson limit.
It is clearly seen in Fig.~\ref{Q_RvsR_M_TOV_andre_p_c_1E18_rho_1E13},
that there are two points where the two lines coincide,
namely, $Q/R=0$ and $Q/R=1$. Indeed, for
$\alpha=0$, i.e. $Q/R=0$, 
both the original interior Schwarzschild limit
and the Buchdahl limit coincide, giving $R/M=9/4$.  For
$\alpha=1$, i.e., $Q/R=1$, both curves give $R/M=
1$, i.e., both limits yield the quasiblack hole solution.  On the
other hand, it is also clearly seen in
Fig.~\ref{Q_RvsR_M_TOV_andre_p_c_1E18_rho_1E13} that although the
lines are near each other, they do not coincide, i.e., the 
electric interior
Schwarzschild and the Buchdahl-Andr\'easson limits are generically
different. Undoubtedly, the values of $R/M$ shown by the solid line are
very close but always larger than those shown by the dashed line. 
This has been substantiated by analytical calculations valid
up to order $Q^2/R^2$
on the interior Schwarzschild limit for these stars, and it
points to the fact that the Buchdahl-Andr\'easson limit is indeed a
limit of limits. As noted in  \cite{andreasson_charged},
the 
$\rho={\rm constant}$ electric solutions 
discussed in 
\cite{giulianirothman}
do not saturate the bound. 
In our work, we also find that our 
$\rho={\rm constant}$ electric solutions,
although close, do not saturate the bound. 
Configurations that do saturate 
the 
Buchdahl-Andr\'easson bound are  self-gravitating electrically 
charged shells \cite{andreasson_charged}. It will be interesting
to investigate whether there are other configurations
that saturate the bound.

\section{Quasiblack hole limit of a charged relativistic 
incompressible star}\label{qbh-section1}

An important point here is to investigate whether one finds
configurations close to the quasiblack hole
configuration or not. 
In the analysis we have found that the limit of the
quasiblack hole configuration is reached for values of the central
pressure much greater than the energy density and for large charge
fractions.

It can be seen from Figs.~\ref{Log_rho_0vsLoR_M_dif_p_c},
\ref{alphavsLoR_M_dif_p_c}, \ref{Log_rho_0vsLoQ_M_dif_p_c},
and \ref{alphavsLoQ_M_dif_p_c} that for some values of the energy density,
central pressure, and charge fraction, the relations $R/M$ and $Q/M$ reach
unity. This limit appears when the central pressure is much larger than
the energy density. In this instance, the parameters that best show these
results are $\rho_0=1.0\times10^{13}$[kg/m$^3$],
$p_c=1.0\times10^{18}$[kg/m$^3$], and $\alpha=0.99$. In this case, 
the quantities
$M$, $R$, and $Q$ are such that $R/M=1.02268$ and $Q/M=0.999853$. These values
can be interpreted as giving $R\simeq M\simeq Q$, indicating that this
charged star is close to the quasiblack hole configuration.

In order to show that there is indeed a quasiblack hole in this 
limit, we analyze also the behavior of the metric functions, $A(r)$ and
$B(r)$, as proposed in \cite{lz1} (see also \cite{lz4,review}).

\begin{figure}[!ht]
     \centering
     \includegraphics[scale=1.145]{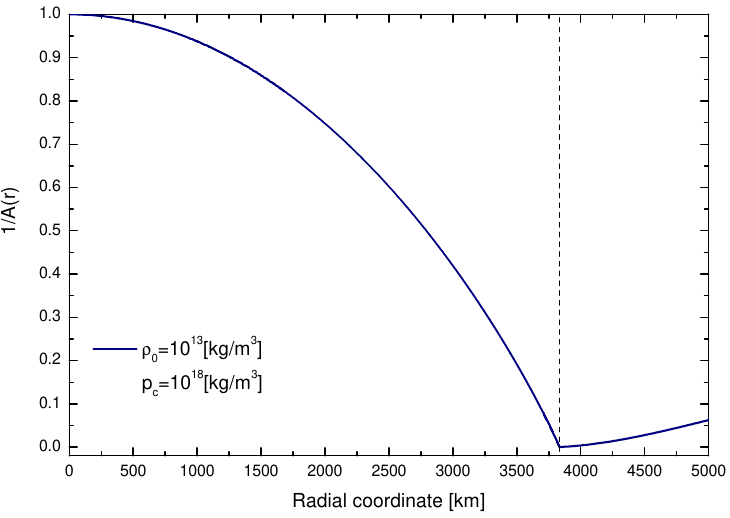}
		\vspace*{-.3cm}
     \caption{The inverse of metric function $A(r)$ as a function of
     the radial coordinate for the quasiblack hole configuration. 
     The parameters used are  $\rho_0=1.0\times10^{13}$[kg/m$^3$],
     $p_c=1.0\times10^{18}$[kg/m$^3$], and charge fraction
     $\alpha=0.99$. The vertical dashed line indicates the surface of
     the star.}
     \label{1_grrvsr_rho_0_1E13_p_c1E18_099}
\end{figure}

We plot $1/A(r)$ versus the radial
coordinate in Fig.~\ref{1_grrvsr_rho_0_1E13_p_c1E18_099} for the
energy density $\rho_0=1.0\times10^{13}$[kg/m$^3$], the central pressure
$p_c=1.0\times10^{18}$[kg/m$^3$], and 
the charge fraction $\alpha = 0.99$. Observe that
the function $1/A(r)$ starts at unity at $r=0$ (no conical
singularity) and decreases with the
increasing of the radial coordinate, reaching its minimum value at the
star's surface $r=R$. The value of $1/A(r)$ at the surface of
the star is approximately zero, 
actually it is $2.25942\times10^{-4}$. This 
vanishingly small value of $1/A(r=R)$ 
indicates that the solution is close to a quasiblack hole
configuration \cite{lz1}.

\begin{figure}[!ht]
     \centering
     \includegraphics[scale=1.145]{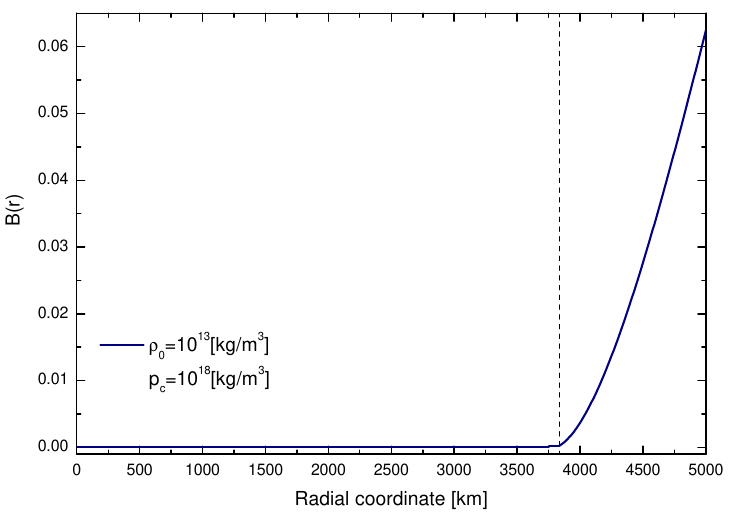}
		\vspace*{-.3cm}
     \caption{The metric function $B(r)$ as a function of
     the radial coordinate for the quasiblack hole configuration. 
     The parameters used are  $\rho_0=1.0\times10^{13}$[kg/m$^3$],
     $p_c=1.0\times10^{18}$[kg/m$^3$], and charge fraction
     $\alpha=0.99$. The vertical dashed line indicates the surface of
     the star.}
     \label{gttvsr_rho_0_1E13_p_c1E18_099}
\end{figure}

We plot the metric potential $B(r)$ as a function of the radial
coordinate in Fig.~\ref{gttvsr_rho_0_1E13_p_c1E18_099} for the same
values of energy density, pressure, and charge fraction as in the case
of the function $1/A(r)$. 
Once the values for the total 
mass $M$, radius $R$ of the star, 
and total charge $Q$ are already known, $B(r)$ is obtained by
numerically integrating the differential equation
\eqref{conservacion2} from the surface to the center of the star. The
results show that the function $B(r)$ assumes values close to zero in
the whole interior region, confirming that this is on the verge of
being a quasiblack hole solution \cite{lz1}. The value of $B(r)$ increases 
albeit slowly with
the radial coordinate, reaching its maximum interior
value at the surface of
the star ($r=R$). This value of $B(R)$ coincides with the minimum
value of the function $A^{-1}(R)$, i.e.,
$B(R)=2.25942\times10^{-4}$. The equality between $B(r)$ and
$A^{-1}(r)$ at $r=R$ satisfies the boundary conditions,
namely, 
$B(R)=A^{-1}(R)=1-\dfrac{2M}{R}+\dfrac{Q^2}{R^2}$. This comes from the
junction conditions which impose continuity of the metric functions.

The event and the Cauchy horizons of a Reissner-Nordstr\"om spacetime
are given by the solutions of the equation $B(r)=0$, i.e.,
$r_\pm\equiv M\pm\sqrt{M^2-Q^2}$, respectively.  Then, since the mass,
the radius, and the charge of the star are very close to each other,
i.e., $R\simeq M\simeq Q$, the event and the Cauchy horizons obey
$r_\pm\simeq M$, with $R\gtrsim r_+$. This indicates that the solution
is a regular static configuration.  Since $R\simeq r_+$, we see that
the boundary of the star approaches its own gravitational radius,
i.e., a quasiblack hole with pressure is close to being reached.  

\begin{figure}[!ht]
     \centering
     \includegraphics[scale=0.29]{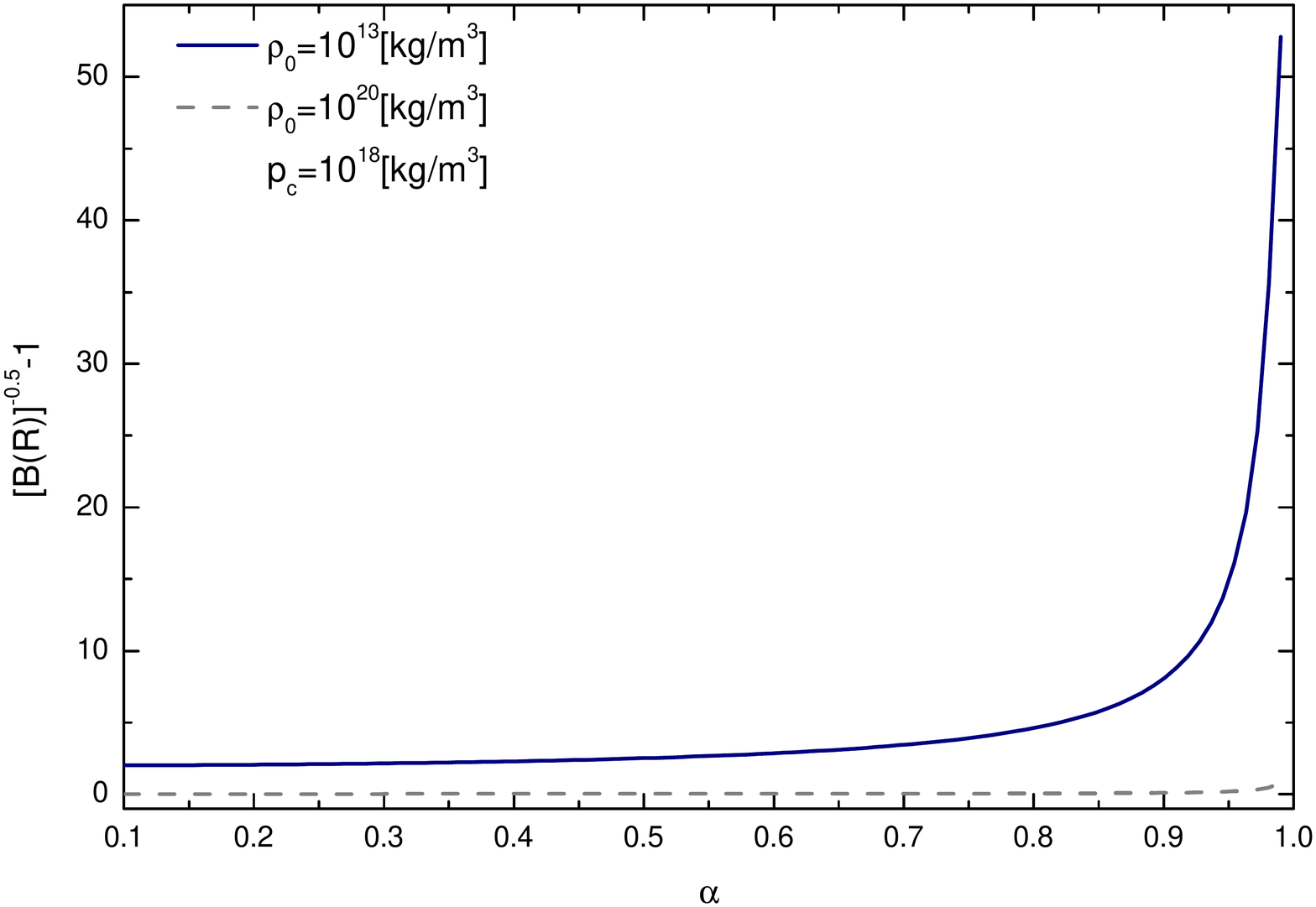}
		\vspace*{-.3cm}
     \caption{The redshift function 
$B(R)^{1/2}-1$ at the surface of the quasiblack hole as a
 function of the charge fraction for the central pressure
 $p_c=1.0\times10^{18}$[kg/m$^3$] and two values of the energy density 
 is displayed. The charged fraction $\alpha$ is varied from
 $0.1$ to $0.99$.}
     \label{redshiftvsalpha1}
\end{figure}

An important quantity is the redshift at the surface of the system,
i.e., how the light frequency of a wave emitted from the system's
surface is redshifted away when it arrives at infinity.  We have
investigated numerically the behavior of the redshift for the charged
incompressible stars presented, comparing the charged and uncharged
cases for a few central pressures and central densities.  The results
are as expected from previous studies \cite{lz1} (see also
\cite{bonnor_wick,lemosweinberg}).  In the quasiblack hole limit, the
redshift at the surface of the stars is indefinitely large.  However,
numerically we were able to find values of the order of 100,
i.e., the larger ratio between the frequency of a flash of light
emitted at the surface of the star compared to the observed frequency
at spatial infinity we found numerically was of about 100 (the
low ratio is due to numerical convergence problems in the limit of
large charge fractions).
In
Fig.~\ref{redshiftvsalpha1}, the redshift function $B(R)^{1/2}-1$ at
the quasiblack hole's surface is displayed for $0<\alpha<1$.

Quasiblack
hole behavior has been also found in
\cite{defelice_siming,defelice_yu,annroth} for an incompressible
fluid, where different equations for the charge density have been
considered. Other works where quasiblack holes have been found are
\cite{lemosezanchin_QBH_pressure,alz-poli-qbh,bonnor_wick,lemosweinberg,bronn},
to name a few.

\section{Conclusions}\label{sec-conclusion}

In this work we have studied stars made of an incompressible fluid,
i.e., a fluid with constant energy density $\rho_0$, and with an
electrical charge distribution $\rho_e$ proportional to $\rho_0$,
$\rho_e=\alpha\,\rho_0$, with $\alpha$ a
number between $0$ and $1$. The analyzed
configurations have spherical symmetry whose exterior spacetime is
described by the Reissner-Nordstr\"om metric.  The configurations were
analyzed for different values of the energy density, central pressure
and charge fraction. 
We have found that the electric interior Schwarzschild limit
does not saturate the 
compactness Buchdahl-Andr\'easson bound, except
in the border solutions $Q/M=0$ and $Q/M=1$. 
For the highest value of the charge fraction that
still yields reliable numerical results, i.e., $\alpha=0.99$, and for
a central pressure much larger than the energy density, we showed that
there is a highly 
compact configuration with $R\simeq r_+$, i.e., a configuration on
the verge of becoming a quasiblack hole.

Our results supplement the results of several
previous works. 
In relation to the Buchdahl-Andr\'easson bound,
as remarked in \cite{andreasson_charged},
the work \cite{giulianirothman}  presents that electric stars
with $\rho=\rho_0$ and $\rho_0$ a constant
do not saturate the Buchdahl-Andr\'easson bound.
Surely, the set of solutions found in 
\cite{defelice_yu,defelice_siming,annroth},  
which include the solution 
discussed in \cite{giulianirothman}, 
does not saturate the Buchdahl-Andr\'easson bound.
In addition, analytical calculations,  valid up 
to order $Q^2/R^2$,
on the stars we have presented here also show that 
the Buchdahl-Andr\'easson bound is not saturated.
In relation to quasiblack holes,
on one hand, our work
shows that the results found in \cite{defelice_yu,defelice_siming,annroth}
are robust.  In \cite{defelice_yu,defelice_siming,annroth} it was 
considered incompressible electrically charged fluids $\rho=\rho_0$
and $\rho_0$ a constant,
but with a charge distribution that differs from ours,
namely, $q(r)=Q\,(r/R)^n$ for some exponent $n\geq3$
in \cite{defelice_yu,defelice_siming}, or a more intricate  
distribution \cite{annroth}. Numerical star
solutions were found and it was shown that configurations with
boundaries as close as one wants to their own gravitational radius are
possible. I.e., quasiblack holes also appear in
\cite{defelice_yu,defelice_siming,annroth}. On the other hand, our
previous work \cite{alz-poli-qbh}
analyzes with care compressible configurations of stars with
an equation of state $p=\omega\,\rho^\gamma$, for some constant $\omega$
and exponent $\gamma$, and also discusses the incompressible
$\gamma\to\infty$ limit. In this work \cite{alz-poli-qbh},
the name Buchdahl limit 
was used for what we have called here more appropriately 
the interior Schwarzschild limit.
Furthermore, the results presented
here are also related to those displayed
in \cite{lemosezanchin_QBH_pressure}
for the Guilfoyle exact configurations \cite{guilfoyle} 
which
have a Cooperstock-de la Cruz-Florides equation of state 
\cite{cooperstock,florides} and a different equation for the charge
distribution. 
In brief, for a wide range of parameters, 
the structure of the charged stars change slightly 
when one changes the equations of state.
However, not 
all equations of state have a quasiblack hole limit.
An example of an equation of state that does not yield 
stars with a quasiblack hole limit is 
a polytropic equation of state, see e.g. 
\cite{raymalheirolemoszanchin} and 
\cite{alz-poli-qbh}.

\begin{acknowledgments}
\noindent 
JPSL and VTZ thank Francisco Lopes and Gon\c calo Quinta for
conversations and collaboration on the analytical calculations for the
interior Schwarzschild limit.  JDVA thanks Coordena\c{c}\~ao de
Aperfei\c{c}oamento de Pessoal de N\'\i vel Superior - CAPES, Brazil,
for a grant.  VTZ would like to thank Conselho Nacional de
Desenvolvimento Cient\'ifico e Tecnol\'ogico - CNPq, Brazil, for
grants, and Funda\c{c}\~ao de Amparo \`a Pesquisa do Estado de S\~ao
Paulo for a grant (Processo 2012/08041-5).  JPSL thanks the Funda\c
c\~ao para a Ci\^encia e a Tecnologia of Portugal - FCT for support,
Projects No.~PTDC/FIS/098962/2008 and
No.~PEst-OE/FIS/UI0099/2011. JPSL and VTZ thank the Observat\'orio
Nacional do Rio de Janeiro - ON, Brazil, for hospitality.

\end{acknowledgments}

\appendix \section{Equation of structure in dimensionless
form}\label{appendixA}

For the numerical calculations it is convenient to write the
equations of structure in a dimensionless form. We then introduce
a rescaled radial coordinate $\varepsilon$ through the equation
\begin{equation}
r=\frac{\varepsilon}{\sqrt{4\pi\rho_0}}\, ,
\end{equation}
where we have made $G=1$ and $c=1$.
In addition, new variables $\mu(\varepsilon)$, $\theta(\varepsilon)$, and
$\kappa(\varepsilon)$ are defined in terms of $m(r)$, $p(r)$, and $q(r)$,
respectively, by
\begin{eqnarray}
m(r)=\frac{\mu(\varepsilon)}{\sqrt{4\pi\rho_0}},\\
p(r)= p_c\,\theta(\varepsilon)\,,\\
q(r)=\frac{\varepsilon^{2}\kappa(\varepsilon)}{\sqrt{4\pi\rho_0}},
\end{eqnarray}
where 
$\rho_0$ and $p_{c}$ represent the energy density and central pressure
of the star, respectively. Now, in terms of the new variables $\mu$,
$\theta$, and $\kappa$, Eqs. (\ref{continuidadedacarga}),
(\ref{continuidaddamassa}), and (\ref{tov}) in dimensionless form become
\begin{eqnarray}
&&\frac{d\kappa}{d\varepsilon}= -\frac{2\kappa}{\varepsilon}+
\dfrac{\alpha}{\sqrt{1-\dfrac{2\mu}{\varepsilon}+
\varepsilon^2 \kappa^2}}\,,\label{kappa}\\
&&\frac{d\mu}{d\varepsilon}=\varepsilon^{2}
+\dfrac{\alpha\varepsilon^3 \kappa}
{\sqrt{1-\dfrac{2\mu}{\varepsilon}+\varepsilon^2\kappa^2}}\,,
\label{mu}\\
& &\frac{d\theta}{d\varepsilon}=
-\varepsilon\left(\theta+\rho_0 p_{c}^{-1}\right)\left(
\dfrac{\displaystyle{p_c\rho_{0}^{-1}\theta -\kappa^2+
\frac{\mu}{\varepsilon^3}}}{1-
\dfrac{2\mu}{\varepsilon} +\varepsilon^2 \kappa^2}\right)\nonumber \\
& &\hskip 1cm+\dfrac{\displaystyle{\alpha \rho_0 p_{c}^{-1}\kappa}} {
\sqrt{1- \dfrac{2\mu}{\varepsilon} +
\varepsilon^2 \kappa^2 \,}}\,,
\label{theta}
\end{eqnarray}
where we have also used relation (\ref{densicarga_densimassa}) written in
the new variables.
This set of coupled differential equations, (\ref{kappa})-(\ref{theta}),
is solved to get the equilibrium solutions. The boundary conditions
adopted in the center of the star, i.e., at $\varepsilon=0$, are
$\kappa(0)=0$, $\mu(0)=0$, and $\theta(0)=1$.

Once the functions $\mu(\varepsilon)$ and $\kappa(\varepsilon)$ have been
obtained by means of numerical integration, the metric potential
$A(\varepsilon)$ is obtained from its definition,
 Eq.~\eqref{funcion metrica},  which in the
dimensionless variables reads
\begin{eqnarray}\label{A1}
& & A^{-1} = 1- \dfrac{2\mu}{\varepsilon} +
\varepsilon^2 \kappa^2\,,
\end{eqnarray}
while the metric potential 
$B(\varepsilon)$ is determined by integration of Eq.~\eqref{conservacion2},
\begin{eqnarray}
 \label{B1}
& & \frac{dB}{d\varepsilon}=2\varepsilon\, B\,\left(
\dfrac{\displaystyle{ p_c\rho_{0}^{-1}\theta-\kappa^2+
\frac{\mu}{\varepsilon^3}}}{1-
\dfrac{2\mu}{\varepsilon} +\varepsilon^2 \kappa^2}\right),
\end{eqnarray}
where again we have replaced the original variables $r$, $q(r)$, $m(r)$,
and $p(r)$, by the respective dimensionless quantities
$\varepsilon$, $\kappa(\varepsilon)$, $\mu(\varepsilon)$, and
$\theta(\varepsilon)$.

The integration of equations \eqref{kappa}-\eqref{theta}
comes to a halt at the point where the pressure $\theta$
reaches zero value, finding thus the value of $\varepsilon$ at the
surface of the star,
$\varepsilon=\varepsilon_s$.  The corresponding value of the radial
coordinate is extracted and the radius of
the sphere is obtained from the relation
$R=\dfrac{\varepsilon_s}{\sqrt{4\pi\rho_0}}$. The other physical
quantities, such as the mass $M$ and the charge $Q$, are
calculated from the relations $M\equiv
m(R)=\dfrac{\mu(\varepsilon_s)}{\sqrt{4\pi\rho_0}}$ and $Q\equiv
q(R)=\dfrac{\varepsilon_s^{2}\kappa(\varepsilon_s)}{\sqrt{4\pi\rho_0}}$,
respectively.

\end{document}